\documentclass[12pt,aps,prb,preprint,showkeys]{revtex4-1}

\usepackage{amsmath}    % need for subequations
\usepackage{graphicx}   % for figures

\newcommand{\conopt}{& Very & Somewhat & Not very & Not at all \\ \hline}
\newcommand{\ucnopt}{& Improved & Stayed the same & Declined \\ \hline}
\newcommand{\intopt}{& 0 & 1 & 2 & 3 \\ \hline}

\begin{document}

\title{Enhancing Introductory Student Motivation with a Major-Managed Course Blog: A Pilot Study}
\author{Ashley A. August}
 \affiliation{Department of Physics, Jacksonville University, Jacksonville, FL 32211}
\author{Kenneth C. Bretey}
 \affiliation{Department of Physics, Jacksonville University, Jacksonville, FL 32211}
\author{Bryant T. Cory}
 \affiliation{Department of Physics, Jacksonville University, Jacksonville, FL 32211}
\author{Elliott R. Finkley III}
 \affiliation{Department of Physics, Jacksonville University, Jacksonville, FL 32211}
\author{Robbie D. Jones}
 \affiliation{Department of Physics, Jacksonville University, Jacksonville, FL 32211}
\author{Dennis W. Marshall}
 \affiliation{Department of Physics, Jacksonville University, Jacksonville, FL 32211}
\author{Phillip C. Rowley}
 \affiliation{Department of Physics, Jacksonville University, Jacksonville, FL 32211}
\author{W. Brian Lane}\email{corresponding author: wlane@ju.edu}\affiliation{Department of Physics, Jacksonville University, Jacksonville, FL 32211}
\date{\today}

\begin{abstract}
Enhancing motivation and learning attitudes in an introductory physics course is an important but difficult task that can be achieved through class blogging. We incorporated into an introductory course a blog operated by upper-level physics students. Using the Colorado Learning Attitudes about Science Survey (CLASS), periodic in-class surveys, analysis of student blog comments, and post-instructional interviews, we evaluate how the blog combined with class instruction provided the students with a better sense of relevance and confidence and outline recommendations for future use of this strategy.
\end{abstract}

\maketitle

\section{INTRODUCTION}
\subsection{The Importance of Motivation in Physics Education}

Physics education research has identified a variety of student learning attitudes that shape and are shaped by the learning experience,\cite{Hammer:00} many of which are not explicitly addressed in typical introductory courses.\cite{Redish}\textsuperscript{,}\cite{Tarshis} Similarly, learning theory describes the importance of considering student motivation in curriculum design.\cite{Ames}\textsuperscript{,}\cite{Keller2} In particular, it is important for science instructors to consider the motivational factors of relevance and confidence,\cite{Hammer:94}\textsuperscript{,}\cite{Redish} which have a significant impact on student performance in both class and lab\cite{Lynch} since they determine the amount and quality of effort that the students put forth.\cite{Burke}

Instruments that measure student learning attitudes and motivational factors reveal that introductory physics courses typically result in notable declines.\cite{CLASS}\textsuperscript{,}\cite{Redish} The challenge for instructors, therefore, is to design instructional strategies that can encourage, develop, and reward these learning attitudes and motivational factors.

\subsection{Course Blogs: A Possible Remedy}

Weblogs can be used in an introductory physics course to improve attitudes and expectations by providing an opportunity for the students to relate to the course material and to develop their confidence. One study\cite{Duda_a}\textsuperscript{,}\cite{Duda_b} that used an extra-credit course blog written by the instructors to supplement class discussion found that students that actively participated in the blog maintained more positive attitudes towards physics (particularly a stronger sense of relevance of physics concepts) throughout the course than those who did not actively participate. A second study\cite{Harrison} that required the students to compose articles on a course blog found that students engaged at a deeper level on the blog than they did in class or traditional homework because of the freedom to pursue their interests and because of their sense of involvement in an online community.

Course blogs offer a number of useful benefits to instruction because of their ability to$\ldots$
\begin{itemize}
\item Encourage a combination ``of solitary thought and social interaction to engage students and reinforce learning,''\cite{Harrison} which can result in improved confidence.
\item Create a community of learning and collaboration that benefits students of all personality types and learning styles.\cite{Harrison}
\item Help students develop critical thinking and reflection skills.\cite{Harrison}
\item Assess and capitalize on students' pre-instructional interests,\cite{Harrison} thereby improving their sense of relevance.
\item Incorporate into the learning community outsiders at different stages along the novice-to-expert journey.\cite{Sprague}\textsuperscript{,}\cite{Daley}\textsuperscript{,}\cite{Dreyfus}
\item Complement and enhance other instructional strategies.\cite{Higdon}
\item Provide students with real-world experience working with Web media.\cite{Harrison}
\end{itemize}

%A course blog can also be structured with many options, including$\ldots$
%\begin{itemize}
%\item The audience to whom the blog is accessible (e.g., the general public, the class participants, and invited guests).
%\item Students' role in the blog (as authors, commentators, or observers).
%\item How the blog is graded (including expectations of participation frequency and post content).
%\item How blog participation impacts the course grade (extra credit, required, or a combination).
%\end{itemize}

Course blogs can also present certain challenges, such as students' participating under the pressures of the last minute or distractions,\cite{Harrison} which can have a detrimental impact on the students' learning experience and motivation. When using a course blog, the instructor may need to make explicit connections in class between the course material and blog content.\cite{Harrison} Doing so can encourage students to participate and successfully develop their sense of relevance.

In Section II, we describe the implementation and mixed-methods assessment strategy of an introductory course blog managed by upper-level physics majors. In Section III, we present the results of the instructional assessment and in Section IV we present a set of recommendations based on that assessment. After wrapping up with a few conclusions in Section V, we present the assessment materials in Appendix A.

\section{THE PRESENT STRATEGY: A MAJOR-MANAGED INTRODUCTORY COURSE BLOG}

We conducted a semester-long implementation of a blog in a spring 2011 introductory physics course for Aviation majors operated by upper-level physics majors in order to improve the learning attitudes and motivational factors of the introductory physics students. The upper-level physics majors were selected as blog authors to function as ``mediators'' of the course material between the instructor and the introductory students, since the upper-level physics majors are presumably found between the extremes of novice and expert.\cite{Dreyfus}\textsuperscript{,}\cite{Daley}

\subsection{The Introductory Students of Interest}
These Aviation majors train to become commercial or military pilots while pursuing an undergraduate business degree. Typically, Aviation Physics is the only lab science course these students take. The course is designed to provide the students with some understanding of how aircraft and aircraft systems physically operate, producing well-rounded pilots who are more aware and more adaptive. As was revealed by student blog comments and post-instructional interviews, many of these students have not had much, if any, exposure to physics in a formal learning environment.

Insight into these Aviation majors' motivation, learning attitudes toward physics, and experience in the Aviation Physics course can be gained by examining the pre- and post-instructional results of the Colorado Learning Attitudes about Science Survey (CLASS), which assesses student epistemological beliefs about physics and learning physics along various categories by reporting a percent favorable score that indicates the students' agreement with expert-like beliefs.\cite{CLASS} In this study, we focus on the percent favorable score of the overall survey and of the categories of Personal Interest, Real World Connection, Applied Conceptual Understanding, and Problem Solving Confidence. The first three of these categories are related to the motivational factor of relevance%: Personal Interest refers to the students' desire to learn about the material outside the classroom or to apply the material when it is not required, Real World Connection is how well students can take the information learned in the course and apply it to the world around them, and Applied Conceptual Understanding is how well students can take the material they have learned and use the same methods and concepts on similarly formatted problems
; the fourth category is related the motivational factor of confidence%: Problem Solving Confidence refers to how much the students believe that they can solve a given problem before it is attempted
.\cite{CLASS}

Figure 1 depicts the CLASS results for the fall 2010 semester (during which the present instructional strategy was not implemented), indicating that the Aviation majors enter the course with novice-like approaches to physics, and that these beliefs do not make strong shifts toward expert-like approaches. In particular, the Real World Connection percent favorable score showed a statistically significant decrease %(defined as the average shift in score being greater than twice the standard error of student shifts in that score)
from pre- to post-instruction of 10\%, illustrating that the course did not positively impact the students' senses of relevance. Such decline is a common occurrence in introductory physics courses.\cite{Redish}\textsuperscript{,}\cite{CLASS} The only percent favorable score to show a statistically significant increase was that for Problem Solving Confidence, likely due to the fact that problem solving practice is heavily emphasized in the Aviation Physics course.

\subsection{Blog Design}
Because of its ability to link to external sources and provide multimedia-based presentation of concepts, the major-managed course blog was conceived as an opportunity to develop the spring 2011 students' sense of relevance. In the first week of running the course blog, the introductory students were asked what benefits they hoped to gain from the course blog; the majority of the students indicated that they would like the blog to serve as a venue for seeing demonstrations of and practicing the types of problems they would be expected to solve in the course, indicating a felt need for a better sense of confidence.

The blog structure was therefore designed by the authors\cite{PEER} to focus on developing the students' motivational factors of relevance and confidence.\cite{Keller1}\textsuperscript{,}\cite{Scherr} The blog articles were categorized into six topic sequences, each of which was authored by one or more upper-level physics majors or the instructor:

\begin{enumerate}
\item \emph{Physics in the News}. This topic sequence showed the introductory students how physics relates to the real world, and how physics is an ever-evolving field of study.
\item \emph{Physics in Aviation}. This topic sequence directly related physics to the Aviation majors' primary field of study.
\item \emph{Physics in Entertainment}. This sequence provided a discussion of physics principles (both accurate and inaccurate) in movies and television shows, which has been demonstrated to be an effective instructional supplement.\cite{Hadz}
\item \emph{Physics Humor}. This topic sequence asked the students to apply the concepts of the course to understand physics-related humor. Such humor-related content has been demonstrated to be an effective instructional supplement.\cite{Tatalovic}
\item \emph{Sample Problems}. Usually, students only encounter sample problems in the textbook (which is archived but not interactive) and in the classroom (which is interactive, but not always well archived); these articles provided an interactive and automatically archived learning experience.
\item \emph{Graphing Project Head Start}. The introductory students were given a weekly assignment that asked them to investigate a theoretical model of interest, involving the creation and discussion of a graph depicting the results of their model. This topic sequence helped them begin these assignments by including tips, reminders of previously learned techniques, and further explanations of the model.
\end{enumerate}

In the established motivational framework,\cite{Keller1}\textsuperscript{,}\cite{Hammer:94}\textsuperscript{,}\cite{Redish} topic sequences 1 through 4 were designed to enhance the introductory students' sense of relevance, while topic sequences 5 and 6 were designed to enhance the introductory students' sense of confidence. Topic sequence 1 was written by the instructor. Topic sequences 2 through 4 were written by one upper-level major each. Topic sequences 5 and 6 were written by two upper-level majors each. Each upper-level major was required to post an article to the blog once a week, following a set of posting guidelines designed by the upper-level majors to ensure professionalism, consistency, effective writing and multimedia use, and engaging the introductory students in the on-line conversation.\cite{PEER} %The instructor and upper-level majors also periodically posted additional ``guest'' articles for topic sequences other than their assigned sequences.

The upper-level physics majors participated as blog authors in partial fulfillment of a senior-level physics seminar course to develop their skills in communication, collaboration, research, and technology. In the five-stage novice-to-expert transition scheme of Dreyfus \& Dreyfus,\cite{Dreyfus}\textsuperscript{,}\cite{Daley} these upper-level physics majors exhibit characteristics of the third stage called ``competent'' (able to plan and adapt) and are advancing to the fourth stage of ``proficient'' (able to see the big picture). Thus, they served as helpful mediators between the instructor (at the fifth stage of ``expert'') and the introductory students (at the first stage of ``novice'').

\subsection{Means of Assessment}
The impact of the major-managed course blog on introductory student motivation was assessed by the following mixed-methods approach:

\begin{enumerate}
\item The pre-to-post-instruction shifts in the introductory students' responses to the CLASS\cite{CLASS} assessed the general impact of the course.
\item Short, periodic in-class surveys were administered throughout the semester to monitor the introductory students' senses of relevance and confidence.\cite{Keller1} These surveys (found in Appendix A) were based on established design principles for course-related self-confidence surveys and interest checklists.\cite{CAT}
\item The frequency and content of the introductory students' comments on the blog were analyzed to further elaborate on the survey data and to explore to what degree they engaged in each of the course blog's topic sequences. %For example, minimal surface-level engagement (as revealed by sporadic, terse, and closed-ended comments) indicates that the student may have connected with the topic but not experienced deep learning (thereby establishing relevance but perhaps not confidence), while a deep substantive level of engagement (as revealed by frequent, lengthy, and conversation-continuing comments) indicates that the student has connected with and possibly learned from the topic (thereby establishing relevance and confidence).
    Duda \& Garrett\cite{Duda_b} identify five dimensions for evaluating student comments on a course blog: (1) student interactivity, (2) students' introduction of new knowledge, (3) students' relating the post to ``the course material, real-life, or other disciplines,'' (4) ``self-disclosure of prior knowledge or admission of learning,'' and (5) expression of fascination or interest. Similarly, Harrison\cite{Harrison} describes quality blog posts as those that have substantial content and that relate to the principles of the course.
\item End-of-semester interviews with the introductory students were conducted to further investigate their responses to the in-class surveys and the CLASS, to explore how those responses were impacted by the course blog, and to receive feedback from the students about the blog.
\end{enumerate}

\section{ASSESSMENT RESULTS}
\subsection{CLASS Shifts}

Figure 2 shows the results of the pre- and post-instructional CLASS, indicating that, on the whole, these students did not exhibit the decline in percent favorable scores that are typical in the literature,\cite{CLASS} a similar feature to those students who participated in the course blog in Duda \& Garrett's study.\cite{Duda_a} In fact, each category (including those not reported here) saw an increase in the average percent favorable score, though not all increases were statistically significant.

The one category score to display a statistically significant shift was Personal Interest percent favorable% (which was accompanied by a very slight decline in percent unfavorable score)
. The Applied Conceptual Understanding percent favorable score showed an increase of comparable size (though not statistically significant)%, and the Applied Conceptual Understanding percent unfavorable score showed only a slight increase
. The Real-World Connection percent favorable %and percent unfavorable
score showed only a slight increase, but when compared with the shifts from the previous semester (which showed a statistically significant decline of 10\%% and an increase of 12.5\%, respectively
), it would seem that the spring 2011 course was successful in at least maintaining the students' strong pre-instructional sense of Real-World Connection. %In other words, at the end of the semester, these students maintained their level of effort to explore how physics applies to the real world.
These results all indicate that, on the whole, the students' motivational factor of relevance seems to have been positively impacted by the course.

Examining the Problem Solving Confidence category, we see that the students' average percent favorable score increased somewhat%, while the percent unfavorable score decreased slightly
. These results are not as strong as in the previous semester, but still represent a positive impact on the students' sense of confidence in applying the course material.

The comparison of the pre- and post-instructional CLASS results indicates that the course made a positive impact on the students' senses of relevance and confidence in the course material. Such impact is notable when compared with results typical in the literature\cite{CLASS} which exhibit a notable decline across CLASS categories. The task that remains is to identify if and how the course blog contributed to these changes; to do so, we turn to the in-class surveys, the blog comments, and the post-instructional interviews.

\subsection{In-Class Surveys}

Short in-class surveys (found in Appendix A) were given every two weeks to determine shifts in the students' motivational factors of relevance and confidence and what they perceived to cause those shifts. In general, students reported improvement in their self-confidence in each of the identified skills. When asked which learning activities helped to improve their confidence, students cited the laboratory exercises and in-class activities (particularly the group problem-solving sessions) as being particularly helpful, with a few of the students citing the course blog as being helpful.

When asked about their change in interest on the various topics of the course, student responses were mixed. Students who did report an increase in topics interest primarily identified the lab and class discussions as leading to their increased interest. When describing their experience in the lab, students indicated that the hands-on nature of lab work was very important to their improved interest. Two of the students credited the blog as improving their interest, though one of those students never commented on the blog, which we will discuss further below).

\subsection{Blog Comment Analysis}

The frequency of student comments on the blog was somewhat minimal. Three of the seven introductory students posted a total of 15 comments. However, one student who cited the blog as contributing to his level of interest on the in-class surveys never commented on the blog. This piece of feedback indicates that the students may have read the blog more frequently than they commented on it. We will see this conclusion confirmed in the student interviews. The introductory students' comments did progress from surface-level to substantive over the course of the semester, eventually demonstrating fluency with terminology, precision, and a willingness to apply the concepts of the course to new situations.

Examining the patterns of the introductory students' blog comments yields a number of interesting observations. First, we find that students commented more frequently on articles written by authors with whom they had established a face-to-face relationship. For example, one of the blog authors also worked as a physics tutor on campus, and spent a few tutoring sessions with one of the introductory students; after this tutoring relationship began, the introductory student began commenting on the tutor's articles. Another blog author saw an increase in comments on her articles after she visited one of the introductory students' class sessions. Finally, we note that the instructor's blog articles received the greatest number of comments. From these observations, we tentatively conclude that a face-to-face relationship between the blog participants and authors can be important in encouraging responses. We will see this conclusion confirmed in the student interviews.

Second, the students seemed to primarily respond to articles that were most directly related to what they were studying in class at the time.

Finally, the topic sequences that received the most number of comments were the Graphing Project Head Start, Physics in the News, and Physics Humor. We will further examine the popularity of these topic sequences in the student interviews.

\subsection{Student Interviews}

The post-instruction student interviews largely confirmed many of the findings discussed above.

When asked about how their self-confidence changed over the semester, students identified that their confidence level increased as they applied themselves to the course assignments. ``After a while,'' one student said, ``I got used to$\ldots$ how [the instructor] wanted the assignments turned in, so the assignments became easier.'' ``At the time,'' said another, ``I thought$\ldots$ `This is a lot of work!' but looking back in retrospect, [the assignments] really increased my confidence level.'' ``My confidence level has increased$\ldots$. As we progressed, I have more of a base, so I'm able to figure out things on my own.''

When asked about how their interest level had changed, most students replied that their interest increased most significantly when they had a sense of the applicability of the material, especially when that applicability extended to aviation. One student reported, ``Physics is pretty interesting. You can apply everything we learn in class to something we do in life.'' Another student said that his interest improved ``when we started working with$\ldots$ all that stuff that had to deal with aviation.'' One student also reported that he was excited to see the principles of mathematics applied to real-world situations, which indicates an important lesson about student motivation: They may not see the relevance of what they learn in a course until they are immersed in a subsequent course. Another student related her sense of interest to a growing sense of intrigue with the course material. One student also indicated that his interest was lowest in topics whose problems required many laborious mechanical steps, of which he gave the unit on vectors as an example.

The student interviews revealed that all of the students read the blog, and that they all read more often than they posted comments. A couple students visited just a few times during the semester, while one student reported visiting the blog daily (even though he did not comment on any of the articles). When asked what prevented them from commenting, many students cited busyness and time constraints. ``I did not give as much time as I wished that I could have to the blog,'' one such student regretted. Another student said, ``I think I probably went to the course blog like three times out of the whole semester. I think I was just caught up in all the other stuff, like the assignments.'' One student even stated that when he was able to peruse the blog, he was usually ``burned out on physics.'' Such comments indicate that forgoing blog participation (and the accompanying extra credit) was a necessary sacrifice as the students sought to manage their time and their cognitive load,\cite{de Jong} similar to behavior observed by Harrison.\cite{Harrison}

Some students indicated that they did prefer to interact on the blog with authors with whom they previously had a face-to-face relationship, even if that relationship consisted of only one encounter. The student who participated in out-of-class tutoring with one of the blog authors confirmed that he visited and commented on the blog thanks to this relationship and that, in fact, the author had encouraged him during tutoring to visit the blog. Another student indicated that she felt more encouraged to comment on articles written by one of the authors who had visited the introductory class, even though that one class visit was their only face-to-face contact. These observations confirm the above hypothesis that the students preferred to interact with blog authors whom they had met in person.

Overall, the introductory students reported a positive experience on the blog when they visited it. One student indicated that he was challenged to think critically, while another indicated that it was nice to discuss and work problems with others, reflecting many of Harrison's observations about the benefits of course blogs.\cite{Harrison} Another student indicated that interacting with ``empathetic'' upper-level students helped to change his perceptions of physicists, an excellent benefit derived from the learning community created by the blog. One student who commented frequently confirmed that he felt that his comments became more sophisticated as the semester progressed, thanks to an increase in comfort and knowledge level. ``As the class went on,'' he said, ``I learned more about physics and got more in depth$\ldots$. [My responses] definitely changed to a more analytical way of thinking.''

It seems the variety of topic sequences was valuable to the students. When asked which topic sequences were most helpful, students identified the Physics Humor, Physics in Entertainment, Graphing Project Head Start, and Sample Problems topic sequences. One student also indicated that he would like to see the Physics in Aviation sequence expanded, particularly to include engineering and design topics.

%Five of the seven introductory students indicated that future students would benefit from blog participation being required; two of these five suggested maintaining an offer of extra credit for participation on the blog beyond the required minimum. Two students suggested not making blog participation a course requirement, explaining that they felt the current course assignments were sufficient. To increase student participation, one student recommended advertising the blog more. Another suggested emphasizing the blog as a place to cover topics not touched on in class.

\section{RECOMMENDATIONS}

Overall, it seems that the students' experience of the blog was positive, even when they did not participate in the online discussion. Many students reported that they read the blog regularly, and that the blog did help their learning motivation, if not as significantly as the class activities, lab, and assignments. They particularly reported many of the benefits described in the literature, including a development of critical thinking, engagement, and reinforced learning,\cite{Harrison} and found the blog to complement other learning activities.\cite{Higdon} The interaction with physics majors also seems to have caused a change in at least one student's perception of physicists. The introductory students seem to have appreciated and benefited from nearly all of the topic sequences.

We particularly noted a correlation between increased student participation and established face-to-face relationships, even if that relationship was a one-time classroom visit. We therefore recommend developing face-to-face relationships between blog authors and participant students near the beginning of the blog implementation. For example, the authors may be able to visit the introductory test students' class meeting to deliver a presentation (perhaps in partial fulfillment of a scientific communications course). An instructor could also encourage blog authors and participants to attend a social event outside of class. The blog authors could also conduct demonstrations in or help proctor the introductory course's lab, help the instructor conduct office hours, or serve as private tutors for the introductory course.

%Many of the introductory students favored requiring blog participation. This participation could be used as a means of conducting Just-in-Time Teaching ``Warm-up Assignments,''\cite{JiTT} conducting exam reviews, or hosting virtual office hours. It would also seem beneficial to award extra credit to additional participation on the blog beyond the required minimum. When introducing blog participation as a new requirement, however, an instructor must be careful to consider what currently existing requirements should be scaled back or removed to maintain an appropriate amount of workload for the students. It is important to avoid the perception that the blog is ``busywork.'' We also saw that, as noted by Harrison,\cite{Harrison} the students' experience of the blog was hindered by last-minute participation under the duress of other deadlines and demands on their cognitive load.\cite{de Jong} It would be worthwhile, therefore, to explore the more general question of how different instructors integrate course blogs into their courses, and how that integration determines the nature of the reward for blog participation.

%It also seems important to advertise the blog by various means (in class, over e-mail, via social networking), especially since the instructional strategy is still new to most students.

As with any instructional strategy, we found it vital to assess the introductory students' needs to determine the focus and direction of the course blog. The pre-instructional CLASS, the periodic in-class surveys, and the initial query of student wishes for the blog were very helpful in fulfilling this task.

%We also recommend the use of blogging Web sites such as blogspot.com or Zoho Writer rather than a learning management system (LMS) discussion board to increase ease of use and blog participation by appealing to typical familiarity with quick easy-to-use technology. When using an LMS, only students registered for the course can access the blog content, which would pose difficulties if outside participation was desired. It is particularly noteworthy that Zoho Writer allows users to import equations from \LaTeX, which may be useful for a physics blog both for the convenience and accuracy of \LaTeX and to provide the upper-level students an opportunity to learn and practice \LaTeX typesetting.

Even though its impact on motivation was not formally assessed, the introductory students did report that the hands-on nature of the lab activities significantly helped their motivation. Similarly, when describing which in-class activities helped their motivation, students cited the hands-on nature of in-class problem sessions as important to them. This common characteristic would suggest that the blog should, likewise, include hands-on features. Such features could include computer simulations or instructions for at-home demos.

As described above, this instructional strategy also offers learning benefits to the upper-level majors, including development of their skills in communication, collaboration, technology, and research. These benefits are briefly explored in the reflective section of the students' account of the implementation\cite{PEER} and warrant formal assessment in future implementations of this strategy.

\section{CONCLUSION}

We implemented a course blog managed by upper-level physics majors in an introductory physics course. We evaluated the impact of the course blog using the CLASS, periodic in-class surveys, the student blog comments, and post-instruction student interviews. The blog articles focused on fostering the motivational factors of relevance and confidence in the introductory students. The pre- to post-instructional CLASS shifts show a positive impact on the students' learning attitudes and motivation, especially in the area of Personal Interest; these results were confirmed by the periodic in-class surveys and the interviews. The blog comments revealed an increase in the depth of student engagement on the blog, even though commenting was infrequent. The interviews revealed the students' positive experience with the blog, and have resulted in a number of recommendations for future use.

\newpage

\appendix

\section{In-Class Surveys}

\underline{Course-Related Self-Confidence Survey}

The table below contains a list of skills that you are expected to develop to succeed at this course. For each skill, please circle your current level of self-confidence in that skill.

\begin{table}[h]
  \begin{center}
  \begin{tabular}{|p{0.5\textwidth} | c | c | c | c |}
  \hline
    1. Read and understand a physics problem. \conopt
    2. Select a physics equation to apply to a problem. \conopt
    3. Use algebra to solve a physics equation for a desired unknown quantity. \conopt
    4. Use multiple physics equations in combination to solve a single problem. \conopt
    5. Apply the concepts of trigonometry to a right triangle. \conopt
    6. Apply the concepts of trigonometry to a vector. \conopt
    7. Work with a vector equation. \conopt
    8. Create a graph to depict the relationship between two physical quantities based on predetermined data. \conopt
    9. Create an electronic graph to depict the relationship between two physical quantities based on an equation. \conopt
    10. Examine and draw conclusions from a graph depicting the relationship between two physical quantities. \conopt
  \end{tabular}
  \end{center}
\end{table}

\newpage

\underline{Topics Interest Survey}

The table below contains a list of topics that we will cover in this course. Please circle the number after each topic below that best represents your level of interest in that topic. The numbers stand for the following responses:\\
0 = No interest at all.\\
1 = Interested in an overview of this topic.\\
2 = Interested in reading about and discussing this topic.\\
3 = Interested in applying ideas about this topic in problems or experiments.

\begin{table}[h]
  \begin{center}
  \begin{tabular}{|p{0.5\textwidth} | c | c | c | c |}
  \hline
    1.	Forces and motion \intopt
    2.	Energy \intopt
    3.	The forces of flight \intopt
    4.	Cruising flight \intopt
    5.	Constant-velocity flight \intopt
    6.	Constant-speed turning flight \intopt
    7.	Constant-velocity gliding \intopt
    8.	Electric and magnetic fields \intopt
    9.	DC circuits \intopt
    10.	Resistors \intopt
    11.	Capacitors \intopt
    12.	Inductors \intopt
    13.	AC circuits \intopt
    14.	Thermodynamics \intopt
  \end{tabular}
  \end{center}
\end{table}

\newpage

\underline{Course-Related Self-Confidence Update}

The table below contains a list of skills that you are expected to develop to succeed at this course. For each skill, please circle how your self-confidence in that skill has changed in the last two weeks of the course.

\begin{table}[h]
  \begin{center}
  \begin{tabular}{|p{0.5\textwidth} | c | c | c | c |}
  \hline
    1. Read and understand a physics problem. \ucnopt
    2. Select a physics equation to apply to a problem. \ucnopt
    3. Use algebra to solve a physics equation for a desired unknown quantity. \ucnopt
    4. Use multiple physics equations in combination to solve a single problem. \ucnopt
    5. Apply the concepts of trigonometry to a right triangle. \ucnopt
    6. Apply the concepts of trigonometry to a vector. \ucnopt
    7. Work with a vector equation. \ucnopt
    8. Create a graph to depict the relationship between two physical quantities based on predetermined data. \ucnopt
    9. Create an electronic graph to depict the relationship between two physical quantities based on an equation. \ucnopt
    10. Examine and draw conclusions from a graph depicting the relationship between two physical quantities. \ucnopt
  \end{tabular}
  \end{center}
\end{table}

\underline{Topics Interest Update}

This week, we discussed the topic of \_\_\_\_. How do you feel your interest in this topic has changed (increased, decreased, or remained the same) since the beginning of the week? Please explain.

Please describe what activities led to your change (or lack of change) in interest in this topic. In particular, please describe the role (if any) that class discussions and the course blog played to change your interest in this topic. (If you feel that they did not contribute to your change in interest, please indicate.)

\begin{acknowledgments}
We thank the Aviation Physic students for their participation in this pilot study. We also thank the 2010-2011 Scholarship of Teaching \& Learning Faculty Learning Community (FLC): Heather Downs, Michelle Edmonds, Thomas Harrison (who graciously recorded the student interviews), Andre Megerdichian, Rob Tudor, Colleen Wilson, and Kathy Ingram (who organized the FLC). This work was supported by the 2010-2011 CTL SoTL Fellowship.
\end{acknowledgments}

\newpage
\section*{Figures}

\begin{figure}[h]
\begin{center}
\includegraphics[width=\textwidth]{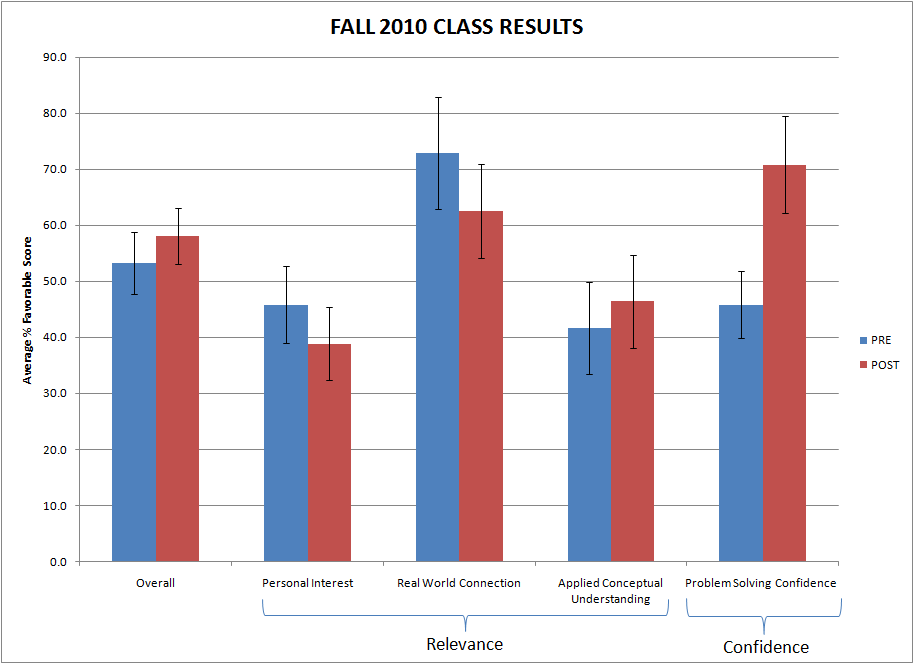}
\caption{\label{fig:fall10}Fall 2010 CLASS results (N = 12 students), exhibiting typical declines in many categories, though a statistically significant increase in Problem Solving Confidence.}
\end{center}
\end{figure}

\begin{figure}[h]
\begin{center}
\includegraphics[width=\textwidth]{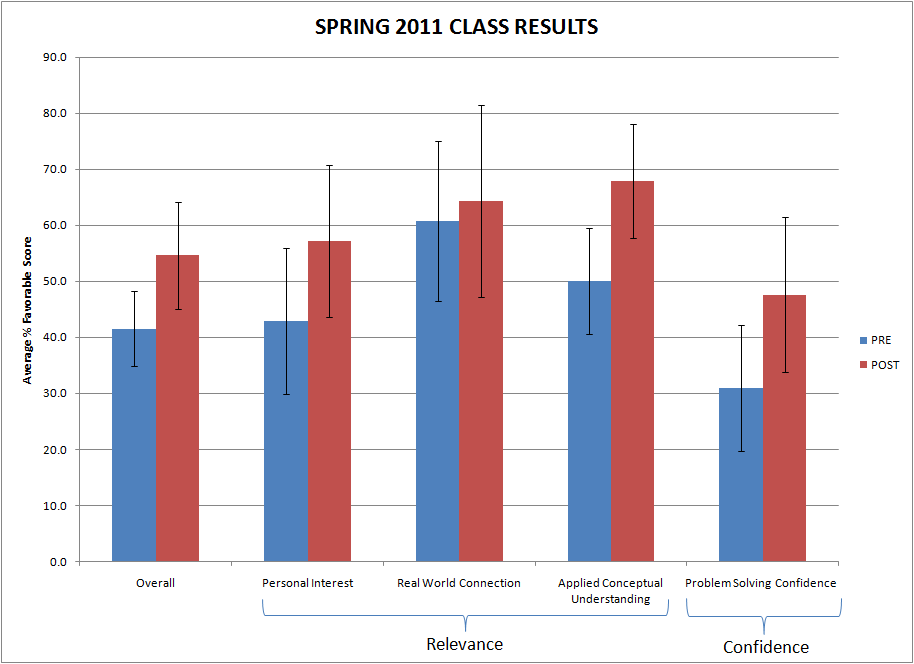}
\caption{\label{fig:spring11}Spring 2011 CLASS results (N = 7 students), without the typical declines.}
\end{center}
\end{figure}

\end{document}